\documentclass{ptephy_arxiv}
\preprintnumber{XXXX-XXXX} 

\usepackage{graphicx}
\usepackage{color}
\usepackage{url}

\begin{document}

\title{Development of a high-light-yield liquid argon detector using tetraphenyl butadiene and silicon photomultiplier  array}


\author{Kazutaka Aoyama}
\author{Masashi Tanaka}
\author{Masato Kimura}
\author{Kohei Yorita}
\affil{Waseda University, 3-4-1, Okubo, Shinjuku, Tokyo, 169-8555, Japan}
\affil{\vspace{-3.5mm}\email{kazutaka-aoyama@kylab.sci.waseda.ac.jp}}
\affil{\vspace{-3.5mm}\email{kohei.yorita@waseda.jp}}



\begin{abstract}%
To increase the light yield of a liquid Ar (LAr) detector, we optimized the evaporation technique of tetraphenyl butadiene (TPB) on the detector surface and tested the operability of a silicon photomultiplier (SiPM), namely, the multi-pixel photon counter with through-silicon-via (TSV--MPPC, Hamamatsu Photonics K.K.) at LAr temperature. 
TPB converts the LAr scintillations (vacuum ultraviolet light) to visible light, which can be detected by high-sensitivity photosensors. 
Because the light yield depends on the deposition mass of TPB on the inner surface of the detector, we constructed a well-controlled TPB evaporator to ensure reproducibility and measured the TPB deposition mass using a quartz crystal microbalance sensor.
After optimizing the deposition mass of TPB (30 $\rm \mu g/cm^2$ on the photosensor window and 40 $\rm \mu g/cm^2$ on the detector wall), the light yield was 12.8 $\rm \pm$ 0.3 p.e./keVee using PMTs with a quantum efficiency of approximately 30\% for TPB-converted light.
We also tested the low-temperature tolerance of TSV--MPPC, which has a high photon-detection efficiency, in the LAr environment.
The TSV--MPPC detected the LAr scintillations converted by TPB with a photon-detection efficiency exceeding 50\%.
\end{abstract}

\subjectindex{}

\maketitle

\section{Introduction}
Owing to its desirable features, liquid argon (LAr) is a widely used scintillator in particle-detection experiments \cite{ref_darkside}\cite{ref_deap}.
For instance, nuclear recoil (NR), and electron recoil (ER) events can be distinguished by their different pulse shapes in the LAr scintillation \cite{ref_LArdecay}.
Pulse-shape discrimination (PSD) significantly reduces the ER background in the search for rare NR events such as dark matter signal.
The ER reduction power of PSD is strengthened by improving the light yield.
Detectors with high-light-yield also benefit the energy reconstruction, especially in regions of low recoil energy.
Therefore, the light yield is a crucial parameter in LAr scintillation detectors.
In this study, we increase the light yield by optimizing the evaporation technique of a wavelength shifter and testing the operability of a silicon photomultiplier (SiPM), namely, the multi-pixel photon counter with through-silicon via (TSV--MPPC, Hamamatsu Photonics K.K.) at LAr temperatures (87 K).

LAr scintillation light is a vacuum ultraviolet that peaks at 128 nm \cite{ref_LArpeak}.
Photosensors, which are operational at LAr temperature, are in general not sensitive to LAr scintillation light.
To eliminate this problem, we convert the LAr scintillation light to visible light (420 nm \cite{ref_TPBpeak1}\cite{ref_TPBpeak2}) using 1,1,4,4-tetraphenyl-1,3-butadiene (TPB) as a wavelength shifter.
The resulting light is detectable by high-sensitivity cryogenic photosensors.
The light yield depends on the deposition mass and surface conditions of the TPB coating on the detector surface, which is deposited using the vacuum evaporation technique \cite{ref_TPBsurface1}\cite{ref_TPBsurface2}.
We thus constructed a well-controlled TPB evaporation system with good reproducibility and measured the deposition mass of TPB using a quartz crystal microbalance (QCM) sensor.
The deposition mass of TPB was optimized by measuring the wavelength-shifting efficiency and transmittance (Section \ref{sec_TPB}).
The light yield is directly proportional to the photon-detection efficiency (PDE) of a photosensor.
Although the sensitivity of TSV--MPPC to visible blue light ($>$50\%) exceeds that of traditionally used PMTs ($<$30\%), the operability of TSV--MPPC at LAr temperature has not been assured.
We therefore checked the low-temperature tolerance of TSV--MPPC in the LAr environment (Section \ref{sec_MPPC}) and demonstrated the operation of a small-size LAr detector with TSV--MPPC arrays (Section \ref{sec_MPPCarray}).
All the measurements were performed at LAr temperature at the test stand in the surface laboratory of Waseda University.
The LAr test stand is described elsewhere \cite{ref_wasedaPMT}.

\section{Optimization of the TPB evaporation technique}\label{sec_TPB}
\subsection{TPB evaporation system}
TPB was evaporated in a stainless vacuum vessel (diameter = 42 cm, height = 40 cm) with a top flange plate to access the internal structures.
The TPB evaporation process was observed through a viewport on the side of the vacuum vessel.
The pressure was reduced to vacuum level of $\rm 1\times 10^{-3}$ Pa using a dry scroll and turbomolecular pumps.
The deposition surface was placed 30 cm above the crucible.
The crucible consisted of a stainless plate containing the TPB, a resistive heater, carbon sheets to improve the thermal uniformity, and a platinum resistance thermometer installed between the stainless plate and heater.
The TPB deposition mass at the height of the deposition surface was measured by QCM sensors (QSET-5P-H, Tamadevice Co., Ltd), which detect nanogram-to-microgram mass changes on the surfaces of the sensor electrodes.
\begin{figure}[ht]
\centering\includegraphics[width=13cm]{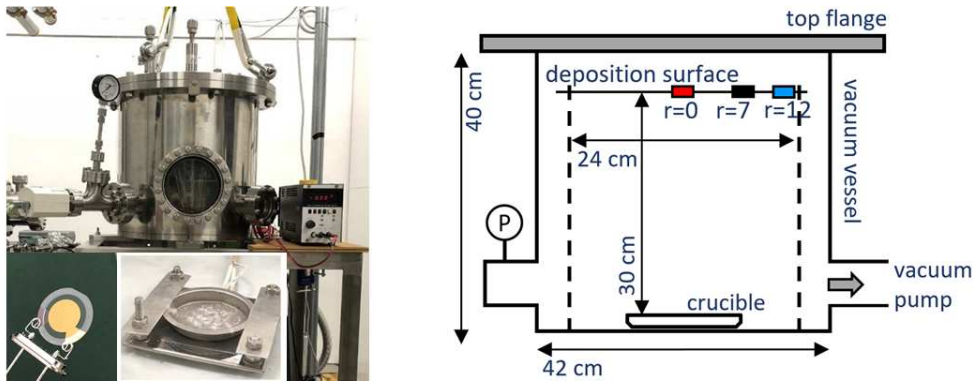}
\caption{
Image (left) and scheme (right) of TPB evaporation system.
The pressure in the vacuum vessel is lower than $\rm 1 \times 10^{-3}$ Pa.
QCM sensors measure the deposition mass of TPB at the height of the deposition surface.
}
\label{fig:Evaporator}
\end{figure}

\begin{figure}[ht]
\centering\includegraphics[width=15cm]{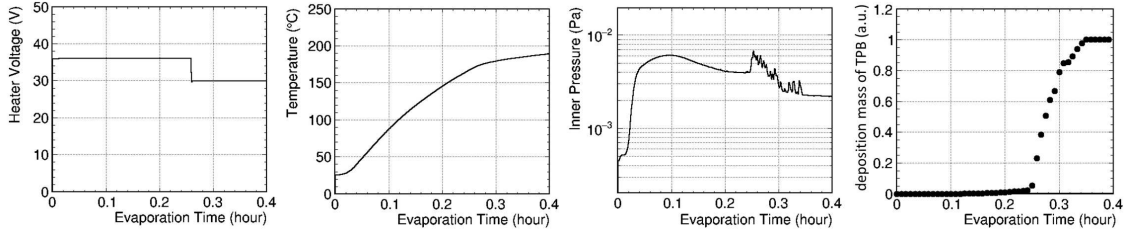}
\caption{
Environmental parameters while heating the TPB: heater voltage (left), crucible temperature (left center), inner pressure (right center), and TPB deposition mass (right).
The horizontal axis indicates the heating time.
}
\label{fig:EvaporationParameters}
\end{figure}

To initiate the evaporation procedure, the vessel was evacuated until its inner pressure reached approximately $\rm 1 \times 10^{-3}$ Pa.
Next, the heater voltage was set to 35 V for the first 15 min, followed by 30 V to prevent TPB degeneration.
The TPB was heated until it was completely evaporated from the crucible.
Fig.~\ref{fig:EvaporationParameters} shows the temporal parameter changes recorded while heating the TPB.
The behaviors of these parameters were reproducible at the same heater voltage, affirming the reproducibility of the TPB coating on the evaporation system.
During the evaporation, we also confirmed the evaporation result by measuring the deposition mass of TPB using the QCM sensor.

\begin{figure}[ht]
\centering\includegraphics[width=6cm]{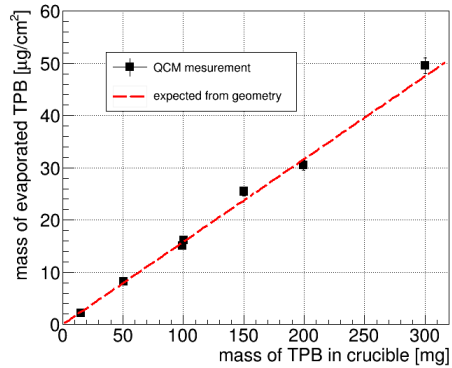}
\caption{
Correlation between the amount of TPB in the crucible and the deposition mass measured using the QCM sensor placed 7 cm away from the center of the evaporation surface.
The black squares and red dashed line plot the deposition masses measured by the sensor and expected from the geometry of the evaporation system, respectively.
}
\label{fig:LinerAndUniform}
\end{figure}

In this system, the TPB deposition mass on the substrate depends on the TPB mass placed in the crucible.
Fig.~\ref{fig:LinerAndUniform} correlates the amount of TPB in the crucible with the deposition mass measured by the QCM sensor placed 7 cm away from the center of the evaporation surface.
The red dashed line in Fig.~\ref{fig:LinerAndUniform} is the TPB deposition mass estimated by assuming that the TPB spreads evenly over the hemisphere from its respective position in the crucible.
The actual deposition mass followed the expected evaporation mass and was a strongly linear function of mass in the crucible.
Moreover, the deposition mass was uniform, varying by less than 10\% within a 7-cm radius from the center axis of the system.

\begin{figure}[ht]
\centering\includegraphics[width=15cm]{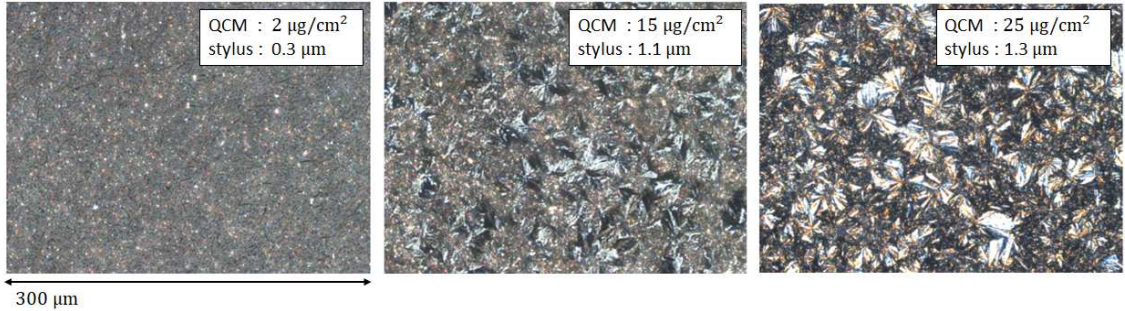}
\caption{
Images of the surface condition of the TPB coating observed under a polarizing microscope.
The crystalline structure grows with increasing TPB deposition mass.
}
\label{fig:EvaporationSurface}
\end{figure}

Fig.~\ref{fig:EvaporationSurface} shows the surface conditions of different TPB deposition masses on acrylic disks.
These images were taken by a polarizing microscope.
The coating with an approximate density of 2 $\rm \mu g/ cm^2$ presented evenly distributed kernels which have relatively high luminance (high polarization).
At higher deposition masses, the coating developed a crystalline structure.
The deposition thickness of TPB was measured using a stylus profilometer (Dektac 6M, ULVAC).
Assuming a close-packed structure with TPB molecular density of 1.079 $\rm g/ cm^3$, the total amount of TPB coating was calculated as 32 $\rm \mu g/cm^2$ (= 0.3 $\rm [ \mu m ] \times 1.079$ $\rm [g/ cm^3 ]$), while the actual TPB mass, measured by QCM sensor, was 2 $\rm \mu g/cm^2$.
Comparing these two values, the filling rate was expected to be approximately 6\%.
The rate got increased with more TPB deposition mass (12\% at 15 $\rm \mu g/ cm^2$ and 18\% at 25 $\rm \mu g/ cm^2$).

\subsection{Optical validation of TPB coating}
\begin{figure}[ht]
\centering\includegraphics[width=11cm]{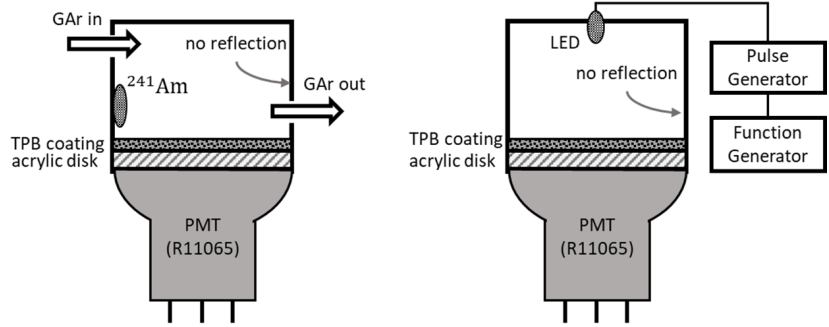}
\caption{
Systems for measuring wavelength-shifting efficiency with the GAr scintillation induced by $\alpha-$particles emitted from $\rm {}^{241}Am$ (left) and transmittance under a blue LED (right).
Measurements were performed on six acrylic disk samples deposited with TPB at different densities (0 to 35 $\rm \mu g/ cm^2$).
}
\label{fig:TPBoptimizeSetup}
\end{figure}

Fig.~\ref{fig:TPBoptimizeSetup} shows the apparatus for measuring the wavelength-shifting efficiency and transmittance of the TPB coating.
The validation samples were six acrylic disks (diameter = 7 cm, thickness = 3 mm) with different TPB densities (0 -- 35 $\rm \mu g/ cm^2$).
A PMT was installed on the non-coated side of each acrylic disk, and a cylinder was installed on the opposite (coated) side of each disk for mounting the light sources.
In the wavelength-shifting efficiency measurement, the gas Ar (GAr) scintillations were induced by $\rm \alpha-$particles emitted from a $\rm {}^{241}Am$ source mounted on the cylinder.
The purity of the GAr was maintained by a continuous flow of GAr into the cylinder.
The PMT detected the converted GAr light passing through the TPB coating and acrylic disk.
In the transmittance measurements, the light source was a pulsed LED with a width of ten ns.
The LED emitted blue light resembling the TPB-converted blue light.

\begin{figure}[ht]
\centering\includegraphics[width=13cm]{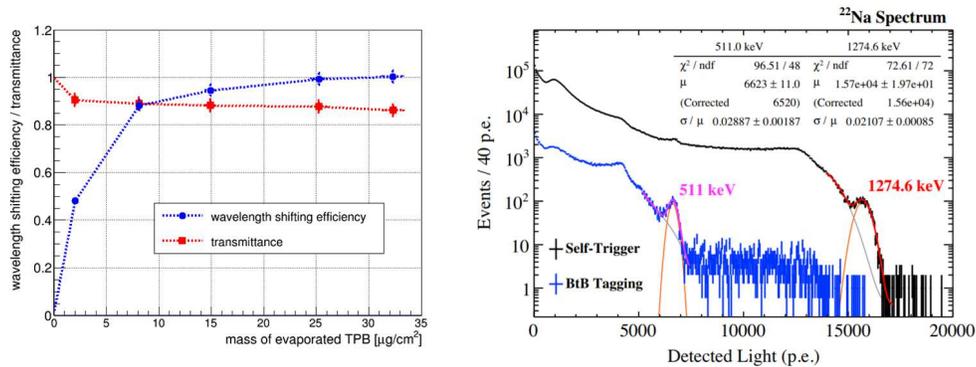}
\caption{
(left) Optical validation results of the TPB coating.
The horizontal axis indicates the density of TPB deposited on the acrylic disks.
(right) Observed spectra of $\rm {}^{22}Na$ detected using the LAr detector, with the adjusted TPB coating based on the validation results.
The light yield of 511-keV gamma rays was 12.8 $\pm$ 0.3 p.e./keVee \cite{ref_wasedaPMT}.
}
\label{fig:resultTPBwt}
\end{figure}

Fig.~\ref{fig:resultTPBwt} shows the measured wavelength-shifting efficiencies (blue circles) and transmittances (red squares) of the six samples.
The wavelength-shifting efficiency was determined as the average number of photoelectrons from the GAr normalized by the highest deposition mass of TPB (the absolute values are not reported).
Meanwhile, the transmittance was determined as the peak PMT signal of the LED light normalized by that of the sample without TPB.
The wavelength-shifting efficiency increased with deposition mass and saturated above 25 $\rm \mu g/ cm^2$.
In contrast, the transmittance decreased with deposition mass but remained around 80\%, even in the saturation region of wavelength-shifting efficiency.

In addition, we measured the light yield using a single-phase LAr detector (the details are provided in \cite{ref_wasedaPMT}).
The LAr detector consisted of a polytetrafluoroethylene sleeve (as the main structure), a reflector foil (3M, enhanced specular reflector (ESR)), and two PMTs (Hamamatsu Photonics K.K., R11065) with a quantum efficiency of approximately 30\% for TPB-converted blue light.
The TPB coating density on the PMT windows was adjusted to the minimum amount (30 $\rm \mu g/ cm^2$) while maximizing wavelength shifting efficiency and to avoid reducing the transmittance of converted light.
The TPB coating density on the ESR was higher (42 $\rm \mu g/ cm^2$) than PMT windows because of the irrelevance of the transmittance.
The right panel of Fig.~\ref{fig:resultTPBwt} presents the spectra of the photons observed with a $\rm {}^{22}Na$ gamma-ray source.
After adjusting the deposition mass of TPB, the light yield of the 511-keV gamma rays emitted from $\rm {}^{22}Na$ was 12.8 $\pm$ 0.3 p.e./keVee.
Considering the scintillation yield of LAr (41 photons/keV \cite{ref_LArabs}), the light yield was limited mainly by the quantum efficiency of PMTs (30\%).
From this measurement, we confirmed the optimality of the TPB evaporation technique.

\section{Low-temperature tolerance test of single-chip TSV-MPPC}\label{sec_MPPC}
\begin{figure}[ht]
\centering\includegraphics[width=12.5cm]{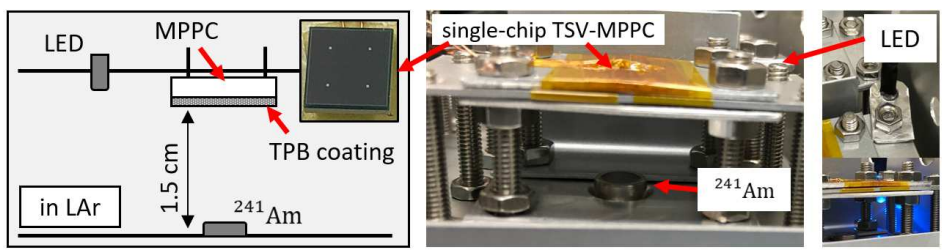}
\caption{
Apparatus of the low-temperature tolerance test.
The single-chip TSV--MPPC was 1.5 cm above the $\rm ^{241}Am$ source of the LAr scintillation light.
The LAr scintillation light was detected through the TSV--MPPC window coated with TPB(30.6 $\rm \mu g/cm^2$).
The average light yield of the LED observed by the MPPC was tuned to less than one photon per pulse.
The apparatus was immersed in LAr during the test.
}
\label{fig:TSV1chsetup}
\end{figure}

Implementing the higher-sensitive photosensor is essential to improve the light yield of the LAr scintillation detector.
Therefore, the operability of the TSV-MPPC, which has a high PDE, was evaluated on the apparatus shown in Fig.~\ref{fig:TSV1chsetup}.
This test was similar to that reported in reference \cite{ref_wasedaVUV}.
The light sources were the LAr scintillations induced by $\rm \alpha$-particles emitted from $\rm ^{241}Am$ (5.5 MeV) and a blue LED with a pulse width of ten ns.
The pulse-generator provided the driving voltage of the LED and the data acquisition trigger for the LED calibration data.
The average light yield of the LED observed by the MPPC was tuned to less than one photon per pulse.
The MPPC in this test was a single-chip TSV--MPPC (Hamamatsu Photonics K.K., S13360-6050VE) with a gain of $1.7 \times 10^6$ at the recommended bias voltage \cite{ref_TSVspec}.
It was $\rm 6\times 6$ mm in size and 50 $\rm \mu m$ pixel pitch.
The TSV--MPPC window (placed 1.5 cm above the $\rm ^{241}Am$ source) was covered by a 30.6 $\rm \mu g/cm^2$ TPB coating.
During the low-temperature tolerance test, the apparatus was all immersed in LAr.
The MPPC driver kit (Hamamatsu Photonics K.K., C12332) supplied the MPPC bias voltage and amplified the signal (the signal amplification gain $G_{amp}$ was 10.9 $\pm$ 0.1).
The MPPC signal was digitized by a flash analog-to-digital converter (Struck, SIS3316) with a 250 MHz sampling rate and was recorded as a waveform.

\begin{figure}[ht]
\centering\includegraphics[width=15cm]{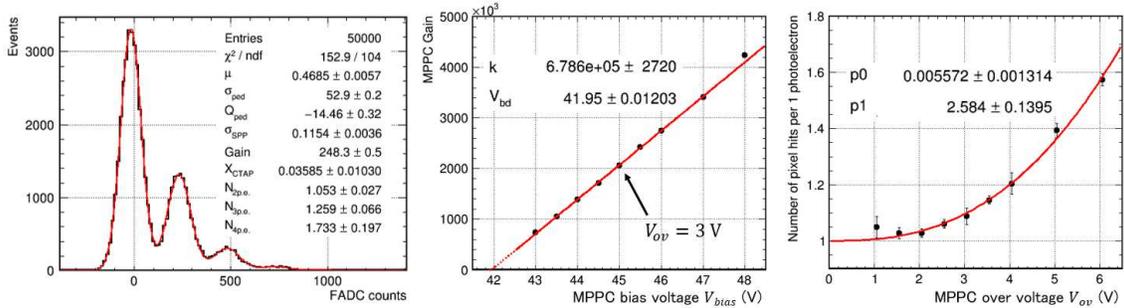}
\caption{
Charge distribution of the LED light (left), and the results of gain (center) and $N_{pix}$ (right) estimated from the charge-distribution fitting.
The red lines are the fitting function.
These parameters indicate that the TSV--MPPC preserves its well-known features even at the LAr temperature (87 K).
}
\label{fig:TSV1chResult}
\end{figure}

The TSV--MPPC was calibrated using the LED data.
The left plot of Fig.~\ref{fig:TSV1chResult} shows the charge distribution obtained at an MPPC bias voltage of 44.5 V.
The signal charge was the integrated value of a waveform over the range [$-$20 ns, +400 ns], based on the timing of the LED pulses.
The first and second peaks in the charge distribution corresponded to pedestal and few-pixel-hits events, respectively.
The MPPC gain was obtained by fitting the charge distribution to the following function:
\begin{eqnarray}
f&=&{\rm Event_{all} \times BinWidth}\times   \\ \nonumber
&\{& {\rm P}(0;\mu) \times {\rm Gaus}(q; Q_{ped},\sigma_{ped})  \\ \nonumber
&+& (1-X){\rm P}(1;\mu) \times {\rm Gaus}(q; Q_{ped}+Q_{Gain},\sqrt{\sigma^2_{ped}+\sigma^2_{SPP}})\  \\ \nonumber
&+& \sum_{n=2}^{} A_n{\rm P}(n;\mu) \times {\rm Gaus}(q; Q_{ped}+n\times Q_{Gain},\sqrt{\sigma^2_{ped}+n\times \sigma^2_{SPP}}) \} \\ \nonumber
\label{eqn:PDEchargeDistLED}
\end{eqnarray}
where ${\rm P}(n;\mu)$ is a Poisson distribution with mean $\mu$, ${\rm Gaus}(q; Q,\sigma)$ is a Gaussian distribution with mean $Q$ and standard deviation $\sigma$, $\rm Event_{all}$ is the total number of events, $\rm BinWidth$ is the bin width of the histogram, $Q_{ped}$ and $\sigma_{ped}$ respectively are a Gaussian mean value and standard deviation of the pedestal, $\sigma_{SPP}$ is standard deviation of the single photon pulse, $Q_{Gain}$ is the difference in charge between adjacent peaks corresponding to the MPPC gain, $X$ is the probability occuring crosstalk and after-pulses, and $A_{n}$ is the scale value of n-th (n $>$ 1) Gaussian.
The fitted parameters were $\mu$, $Q_{ped}$, $\sigma_{ped}$, $\sigma_{SPP}$, $Q_{Gain}$, $X$, and $A_{n}$.
The center and right plots of Fig.~\ref{fig:TSV1chResult} show the fitting results of the MPPC gain ($G_{MPPC}$) and the number of pixel hits per photoelectron ($N_{pix}$), respectively.
The gain was calculated as $G_{MPPC}=Q_{Gain}/(e\times G_{amp})$ at nine bias voltage ($V_{bias}$) points, where $e$ is the elementary charge.
The red line was fitted to the linear function $G_{MPPC}=k\times V_{ov}=k\times (V_{bias}-V_{bd})$ where $V_{ov}$ and $V_{bd}$ are the overvoltages and breakdown voltage, respectively.
We confirmed that the gain increased linearly with bias voltage, and was roughly consistent with the specification value at $V_{ov}=3$ V ($G_{MPPC}=(2.04\pm0.02)\times 10^6$).

The output signal corresponding to one photon from MPPC exceeded the gain because it was interfered with crosstalk and after-pulses.
Therefore, the MPPC signal charge must be corrected for estimating the observed number of photons.
The correction factor is $N_{pix}$, determined as $N_{pix}=Q_{LED}/(Q_{Gain}\times \mu)$, where $Q_{LED}$ is the average charge obtained from the LED light data.
By definition, $N_{pix}$ corresponds to the number of pixel hits per single-photon detection.
In the right plot of Fig.~\ref{fig:TSV1chResult}, the $N_{pix}$ was determined at each overvoltage.
Note that $N_{pix}$ was a nonlinearly increasing function of MPPC overvoltage and was well fitted to $f=1+p_0\times V_{ov}^{p_1}$ (red curve).

\begin{figure}[ht]
\centering\includegraphics[width=15cm]{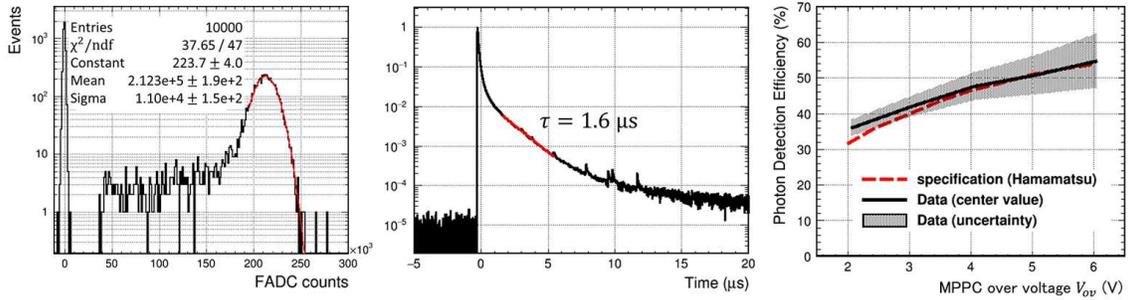}
\caption{
Charge distribution detected by TSV--MPPC, showing a peak of $\rm \alpha$-particles emitted from $\rm ^{241}Am$ (left), average waveform of the LAr scintillation light detected by TSV--MPPC (center), and measured PDE of TSV--MPPC as a function of MPPC overvoltage (right).
The black solid line and red dashed line in the right panel plot the measured PDE and the PDE in the TSV--MPPC specifications \cite{ref_TSVspec}, respectively, and the black band shows the uncertainties in the measurements.
}
\label{fig:TSV1chAmResult}
\end{figure}

The left and center panels of Fig.~\ref{fig:TSV1chAmResult} display the charge distribution of the $\rm ^{241}Am$ data and the average waveform of the events around the peak of the $\rm ^{241}Am$ $\rm \alpha$-particles, respectively.
The lifetime of the average waveform was consistent with the slow component of LAr scintillations ($\tau_{slow}$ = 1.6 $\rm \mu s$ \cite{ref_LArdecay}).
The signal charge was obtained by integrating the waveform over the temporal range t = [$-$1 $\rm \mu s$, +20 $\rm \mu s$].
The PDE was calculated as $PDE = Q_{Am}/(Q_{Gain}\times N_{pix}\times N_{exp})$ where $Q_{Am}$ is the mean charge obtained by fitting the $\rm ^{241}Am$ peak with a single Gaussian function, the gain and $N_{pix}$ were respectively obtained as the fitting results of the center and right plots of Fig.~\ref{fig:TSV1chResult}, and $N_{exp}$ is the expected number of photons reaching the MPPC surface.
$N_{exp}$ was calculated from the following assumptions; the distance between the MPPC surface and the $\rm ^{241}Am$ source (1.5 $\pm$ 0.05 cm), the LAr scintillation emission yield for $\rm \alpha$ particles (27.5 eV/photon \cite{ref_alphaLY}), and the wavelength-shifting efficiency of the TPB coating (100\%).
It was further assumed that photon emission from TPB was uniformly isotropic and that the MPPC detected half the photons converted by TPB.
The right panel of Fig.~\ref{fig:TSV1chAmResult} plots the PDE of the TPB-converted LAr scintillation light as a function of MPPC overvoltage.
Uncertainty in the PDE was due to fitting errors in gain, $N_{pix}$, and $Q_{Am}$, and by misalignment of the distance between the $\rm ^{241}Am$ and the MPPC surface.
The PDE of the TSV--MPPC was approximately 40\% and 50\% at $V_{ov}$ $=$ 3 and 5 V, respectively.
The results confirmed that the TSV--MPPC operates in the LAr environment and detects the TPB-converted visible light with higher efficiency compared to PMTs.

\section{Demonstration of LAr detector using TSV--MPPC Arrays}\label{sec_MPPCarray}
\subsection{Readout Printed Circuit Board of TSV--MPPC array}
\begin{figure}[ht]
\centering\includegraphics[width=14.5cm]{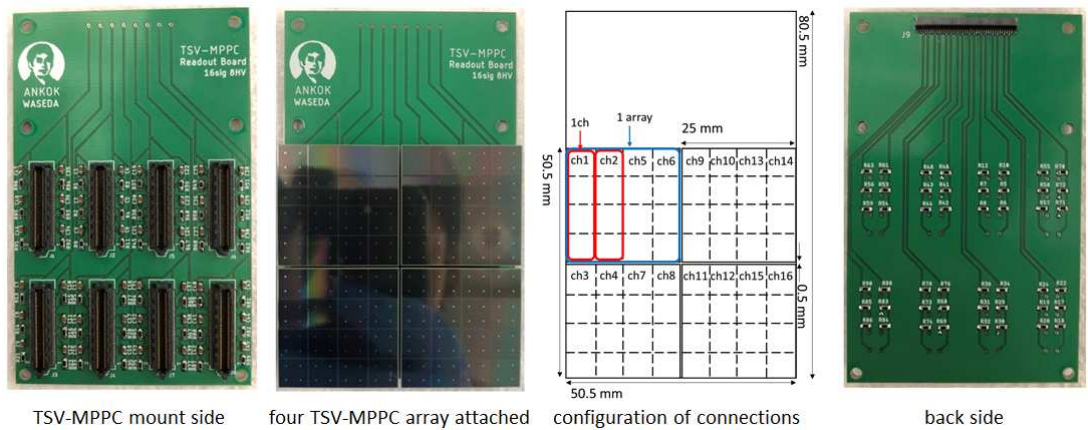}
\caption{
Readout PCB of TSV--MPPC array with the connection circuit.
Four arrays can be mounted on one PCB.
}
\label{fig:ReadoutBoard}
\end{figure}

\begin{figure}[ht]
\centering\includegraphics[width=13.0cm]{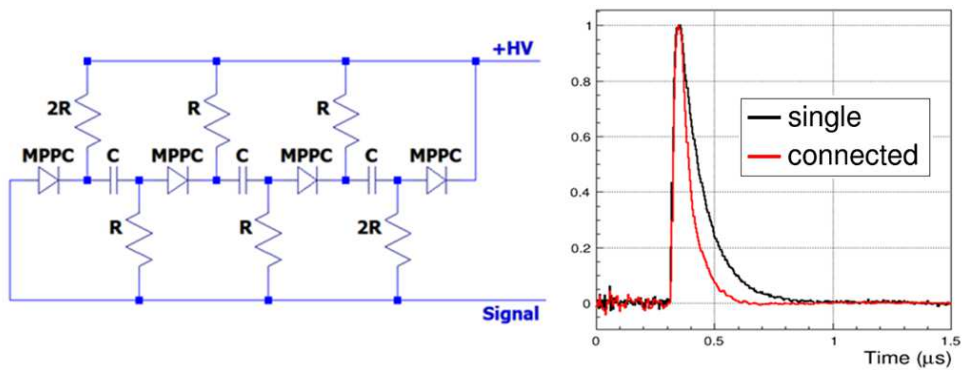}
\caption{
Schematic of MPPC connection \cite{ref_MEG} (left) and average MPPC waveforms of a single-chip (black) and connected chips (red) obtained at room temperature (right).
}
\label{fig:TSVreadoutScheme}
\end{figure}

The basic characteristics were measured using single-chip TSV-MPPC as shown in section \ref{sec_MPPC}, moreover, we demonstrated the operability of the LAr detector using $4 \times 4$ array-type TSV-MPPC because of the necessity of a larger light-detection coverage for practical usage.
The TSV--MPPC array using this study was a $4 \times 4$ assemblage of TSV--MPPC chips (Hamamatsu Photonics K.K., S13361-60**AE-04, ** is the pixel size ("50" or "75") \cite{ref_TSVARRAYspec}), which provides a sufficient light-detection area.
The array was $\rm 2.5\times 2.5$ $\rm cm^2$ in area and the bottom of the sensor was equipped with two electrical connectors.
As shown in Fig.~\ref{fig:ReadoutBoard}, the readout board was a double-sided printed circuit board (PCB) with dimensions of 50.5 mm $\times$ 80.5 mm $\times$ 1.6 mm.
Four (2-column$\rm \times$2-row) TSV--MPPC arrays were installed on the PCB and the connectors (SAMTEC, ST4-40-1.00-L-D-P-TR) were spaced by 0.5 mm to avoid inter-connector interferences at LAr temperature.
To avoid proliferating the number of readout channels, four independent MPPC chips (1-column$\times$4-row) were merged by a connection circuit configured on the readout PCB.
The left panel of Fig.~\ref{fig:TSVreadoutScheme} shows the connection circuit developed in the $\rm \mu \rightarrow e\gamma$ (MEG) experiment \cite{ref_MEG}.
Coupling capacitors (100 nF) were interposed between neighboring MPPCs, and resistances (1.0 $\rm k\Omega$) were connected perpendicular to the line of connected MPPCs and capacitors.
As the MPPC signal was transmitted through the series connection, its lifetime decreased with increasing number of connections.
In contrast, the MPPC bias voltage was supplied as a parallel connection, so was independent of number of MPPC connections.
The right panel of Fig.~\ref{fig:TSVreadoutScheme} compares the average waveforms of the signals obtained from single-chip and connected TSV--MPPCs, respectively, by laser light at room temperature.
The lifetime of the signal was shorter in the connected TSV--MPPC than in the single-chip TSV--MPPC.
Moreover, as two high voltage (HV) lines were merged into one line at the readout PCB, one HV line operated two signal channels (i.e., 8 MPPC chips).
In this configuration, 64 TSV--MPPC chips were driven by eight HV lines and 16 signal lines were extracted from the PCB through a flexible flat cable (FFC).

\subsection{LAr scintillation detector using TSV--MPPC arrays}
\begin{figure}[ht]
\centering\includegraphics[width=15cm]{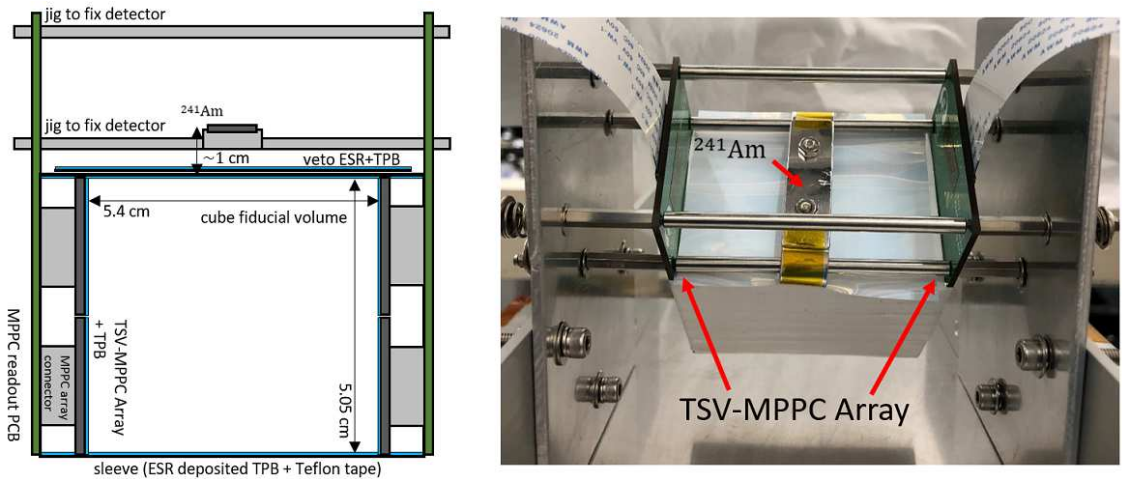}
\caption{
LAr scintillation detector with TSV--MPPC arrays.
The fiducial volume (ESR and Teflon tape) is enclosed within sides of 5 cm.
Eight TSV--MPPC arrays (two readout PCBs) are installed on the two facing sides.
TPB is coated on the ESR (40 $\rm \mu g/cm^2$) and the windows of MPPC arrays (30 $\rm \mu g/cm^2$).
}
\label{fig:detectorMPPC}
\end{figure}

Fig.~\ref{fig:detectorMPPC} shows the single-phase LAr scintillation detector constructed by TSV--MPPC arrays.
A cube fiducial volume, with 5-cm edges, composed of the volumes of the ESR and Telon tape  was installed with eight TSV--MPPC arrays (two readout PCBs, 32 signal channels) on its two facing sides.
To examine the effect of varying the pixel size, we used four MPPC arrays with a pixel pitch of 50 $\rm \mu m$ on one side and four MPPC arrays with a pixel pitch of 75 $\rm \mu m$ on the other side.
The ESR (40 $\rm \mu g/cm^2$) and the windows of the MPPC arrays (30 $\rm \mu g/cm^2$) were coated with TPB.
Gamma rays (59.5 keV) were emitted from a $\rm ^{241}Am$ source installed above the MPPC detector.
The MPPC bias voltage was supplied by an MPPC driver kit (Hamamatsu Photonics K.K., C12332).
To reduce the numbers of feedthroughs and cables, the HV line was divided into eight sublines within the LAr vessel.
The signal from the MPPC exited the LAr vessel through a commercial D-sub feedthrough.
The MPPC signal was passed through a high-pass filter at room temperature and digitized by the flash ADC (Struck SIS3316).

\begin{figure}[ht]
\centering\includegraphics[width=15cm]{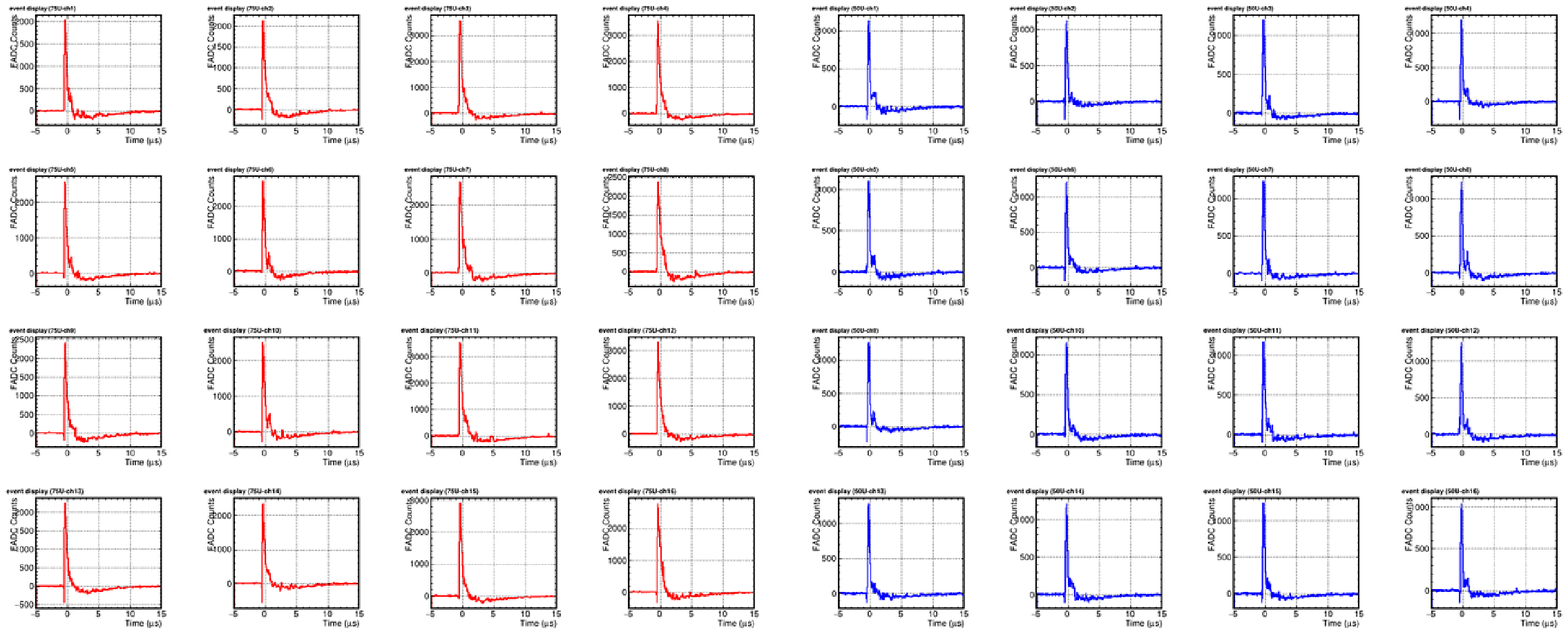}
\caption{
Waveforms of LAr scintillation light detected by TSV--MPPC arrays in the LAr environment.
The red and blue waveforms were obtained at different pixel pitches of the MPPC (75U and 50U, respectively).
The LAr scintillation signals were read from all channels in the detector.
}
\label{fig:WaveformMPPCdetector}
\end{figure}

\begin{figure}[ht]
\centering\includegraphics[width=15cm]{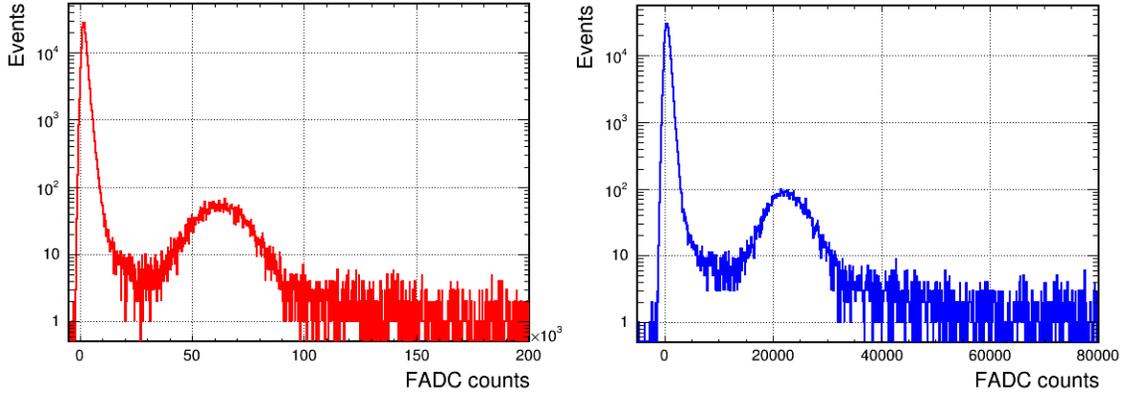}
\caption{
Charge distribution of $\rm ^{241}Am$ obtained by integrating the waveforms at 75U (left) and 50U (right).
The gamma-ray events obviously peak at 59.5 keV.
}
\label{fig:AmDist}
\end{figure}

Fig.~\ref{fig:WaveformMPPCdetector} shows the waveforms of the LAr scintillation events detected by each channel.
All 32 channels detected the LAr scintillation light, the waveforms of the MPPCs with pixel pitches of 75 and 50 $\rm \mu m$ (hereafter referred as to 75U and 50U) are displayed in red (left) and blue (right), respectively.
The gamma-ray emitted from $\rm ^{241}Am$ (59.5 keV) events peaked by integrating the waveform over the temporal range [$-$20 ns, +1 $\rm \mu s$] based on the photon arrival times as shown in Fig.~\ref{fig:AmDist}.
The charges were summed over the 75U or 50U channels without any scaling.

\subsection{Understanding waveforms of events caused by $\rm ^{241}Am$ gamma rays}
\begin{figure}[ht]
\centering\includegraphics[width=15cm]{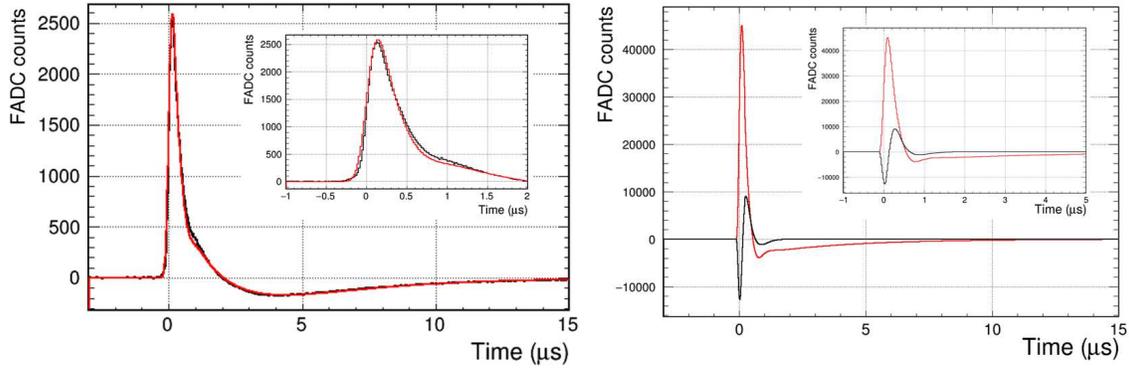}
\caption{
(left)Averaged waveforms around $\rm ^{241}Am$ gamma-ray peak obtained by summing the 75U channels.
The red curves are the fitting results to the MPPC signal templates and the LAr template.
(right) Waveforms of the MPPC signals and electrical cross-talks obtained using pulsed laser light: 75U MPPC signal (red) and electrical crosstalk from 75U MPPCs to 50U channels (black).
}
\label{fig:FittedLAr}
\end{figure}

The left panel of Fig.~\ref{fig:FittedLAr} shows the average waveforms of the events around $\rm ^{241}Am$ gamma-ray peak for the 75U channels.
The waveforms were affected by undershooting and electrical crosstalk caused by the HV connections and D-sub feedthroughs.
Therefore, to explain the signal waveform obtained by the detector, we also fitted the waveform to the convolution of the waveforms of the LAr signal template and the MPPC signal template.
MPPC templates were obtained using a laser with a pulse width of several ns.
The right panel of Fig.~\ref{fig:FittedLAr} shows template waveforms corresponding to 75U MPPC.
The LAr template waveform was obtained by convolving a Gaussian function and two exponential functions with different time constants (a LAr slow component with a time constant of 1.6 $\rm \mu s$ and a LAr fast component with a time constant of 7 ns \cite{ref_LArdecay}).
The ratio of exponential functions with the slow and fast components is a fitting parameter corresponding to the PSD.
Other fitting parameters are the photon arriving time $t_0$, and the signal scales of 75U and 50U.
The fitting was simultaneously applied to the integral waveforms at the 75U and 50U channels.
After fitting, the PSD value (0.68 $\pm$ 0.006) was consistent with that of the ER events detected using the PMT \cite{ref_wasedaPMT}.

\section{Conclusions}
The number of photons observed at a LAr detector depends on the wavelength-shifting efficiency of TPB and the PDE of the photosensors.
We optimized the TPB evaporation technique and evaluated the low-temperature tolerance of TSV--MPPC with high PDE.
The TPB coating was deposited on the substrate using a well-controlled vacuum evaporation system.
We achieved high light-collection efficiency (12.8 $\pm$ 0.3 p.e./keVee) from a PMT--LAr detector with the optimized TPB coating on the inner surface (PMT windows: 30 $\rm \mu g / cm^2$, another surface: 40 $\rm \mu g / cm^2$).
We also confirmed that at LAr temperature, the TSV--MPPC operated with higher detection efficiency ($>$50\%) compared to PMTs ($<$30\%).
The TSV--MPPCs in the LAr detector successfully detected LAr scintillation events.
Implementing the TSV--MPPCs can increase the light yield to approximately 20 p.e./keVee (12.8 $\rm p.e./keVee \times 50\% / 30\%$).
In future work, we will solve the electrical crosstalk and undershoot problems, and hence discuss the absolute values of the photons and the PSD power observed by the TSV--MPPC LAr detector.

\section*{Acknowledgment}
This work is part of the research outcomes of the Waseda University Research Institute for Science and Engineering (project number 2016A-507), supported by JSPS Grant-in-Aid for Scientific Research on Innovative Areas (15H01038/17H05204), Grant-in-Aid for Scientific Research(B) (18H01234), and Grant-in-Aid for JSPS Research Fellow (20J20839).
The authors thank the Material Characterization Central Laboratory at Waseda University for granting access to their stylus profiler.
The authors acknowledge the support of the Institute for Advanced Theoretical and Experimental Physics, Waseda University.


%

\vspace{0.2cm}
\noindent

\let\doi\relax


\begin{thebibliography}{9}

\bibitem{ref_darkside}
\begin{flushleft}
P.~Agnes et al. [DarkSide Collaboration], Phys. Rev. D {\bf 98}, 102006 (2018).
\end{flushleft}
\doi{https://doi.org/10.1103/PhysRevD.98.102006}

\bibitem{ref_deap}
\begin{flushleft}
R.~Ajaj et al. [DEAP Collaboration], Phys. Rev. D {\bf 100}, 022004 (2018)
\end{flushleft}
\doi{https://doi.org/10.1103/PhysRevD.100.022004}

\bibitem{ref_LArdecay}
\begin{flushleft}
Akira Hitachi, Tan Takahashi, Nobutaka Funayama, Kimiaki Masuda, Jun Kikuchi, and Tadayoshi Doke, Phys. Rev. B {\bf 27}, 5279 (1983)
\end{flushleft}
\doi{https://doi.org/10.1103/PhysRevB.27.5279}

\bibitem{ref_LArpeak}
\begin{flushleft}
T. Heindl, T. Dandl, M. Hofmann, R. Kr\"{u}cken, L. Oberauer, W. Potzel, J. Wieser and A. Ulrich, 2010 EPL {\bf 91} 62002
\end{flushleft}
\doi{https://doi.org/10.1209/0295-5075/91/62002}

\bibitem{ref_TPBpeak1}
\begin{flushleft}
W. M. Burton and B. A. Powell, Appl. Opt. 12, 87-89 (1973)
\end{flushleft}
\doi{https://doi.org/10.1364/AO.12.000087}

\bibitem{ref_TPBpeak2}
\begin{flushleft}
Benson, C., Orebi Gann, G.D. \& Gehman, V., Eur. Phys. J. C {\bf 78}, 329 (2018)
\end{flushleft}
\doi{https://doi.org/10.1140/epjc/s10052-018-5807-z}

\bibitem{ref_TPBsurface1}
\begin{flushleft}
R Francini et al, 2013 JINST {\bf 8} P09006
\end{flushleft}
\doi{https://doi.org/10.1088/1748-0221/8/09/P09006}

\bibitem{ref_TPBsurface2}
\begin{flushleft}
V Boccone  et al [The ArDM Collaboration], 2009 JINST 4 P06001
\end{flushleft}
\doi{https://doi.org/10.1088/1748-0221/4/06/P06001}

\bibitem{ref_wasedaPMT}
\begin{flushleft}
M. Kimura, K. Aoyama, M. Tanaka, and K. Yorita, Phys. Rev. D {\bf 102}, 092008 (2020)
\end{flushleft}
\doi{https://doi.org/10.1103/PhysRevD.102.092008}


\bibitem{ref_LArabs}
\begin{flushleft}
Tadayoshi Doke, Akira Hitachi, Jun Kikuchi, Kimiaki Masuda, Hiroyuki Okada and Eido Shibamura, Jpn. J. Appl. Phys {\bf 41} 1538 (2002)
\end{flushleft}
\doi{https://doi.org/10.1143/JJAP.41.1538}

\bibitem{ref_wasedaVUV}
\begin{flushleft}
T.Igarashi, M.Tanaka, T.Washimi, K.Yorita, Nucl. Instrum. Methods Phys. Res., Sect. A 833 (2016) 239-244
\end{flushleft}
\doi{https://doi.org/10.1016/j.nima.2016.07.008}

\bibitem{ref_TSVspec}
\begin{flushleft}
Hamamatsu Corporation. Datasheet S13360 series, 2016 \url{https://www.hamamatsu.com/resources/pdf/ssd/s13360-2050ve_etc_kapd1053e.pdf}
\end{flushleft}

\bibitem{ref_alphaLY}
\begin{flushleft}
Tadayoshi Doke, Kimiaki Masuda, Nucl. Instrum. Methods Phys. Res., Sect. A 420 (1999) 62-80
\end{flushleft}
\doi{https://doi.org/10.1016/S0168-9002(98)00933-4}

\bibitem{ref_TSVARRAYspec}
\begin{flushleft}
Hamamatsu Corporation. Datasheet S13361 series, 2020 \url{https://www.hamamatsu.com/resources/pdf/ssd/s13361-6050_series_kapd1056e.pdf}
\end{flushleft}

\bibitem{ref_MEG}
\begin{flushleft}
K. Ieki et al., Nucl. Instrum. Methods Phys. Res., Sect. A 925 (2019) 148-155
\end{flushleft}
\doi{https://doi.org/10.1016/j.nima.2019.02.010}






\end{thebibliography}






\end{document}